\documentclass[sigconf]{acmart}

\usepackage{graphicx}
\usepackage{balance}  
\usepackage{url}
\usepackage{color}
\usepackage{paralist}
\usepackage{fancyvrb}
\usepackage{epsfig}
\usepackage{subfig}
\usepackage{amsmath}
\usepackage{algorithmicx}
\usepackage{algorithm}
\usepackage[noend]{algpseudocode}
\usepackage{booktabs} 
\usepackage{makecell}
\usepackage{multicol}
\usepackage{listings}

\algrenewcommand\alglinenumber[1]{\scriptsize #1:}
\long\def\comment#1{}   
\newcommand{\q}[1]{\lq\lq{}{}#1\rq\rq{}{}}

\copyrightyear{2019} 
\acmYear{2019} 
\acmConference[SAC '19]{The 34th ACM/SIGAPP Symposium on Applied Computing}{April 8--12, 2019}{Limassol, Cyprus}
\acmBooktitle{The 34th ACM/SIGAPP Symposium on Applied Computing (SAC '19), April 8--12, 2019, Limassol, Cyprus}

\begin{document}
\title[Answering SPARQL Queries through Zero-Knowledge Link Traversal]{How Many and What Types of SPARQL Queries can be Answered through Zero-Knowledge Link Traversal?}

\author{Pavlos Fafalios}
\orcid{0000-0003-2788-526X}
\affiliation{
  \institution{L3S Research Center, Leibniz University of Hanover}
  \city{Hannover, Germany} 
}
\email{fafalios@L3S.de}

\author{Yannis Tzitzikas}
\affiliation{
  \institution{Computer Science Department, University of Crete, and Information Systems Laboratory, FORTH-ICS}
  \city{Heraklion, Greece} 
}
\email{tzitzik@ics.forth.gr}

\begin{abstract}
The current de-facto way to query the Web of Data is through the SPARQL protocol, where a client sends queries to a server through a SPARQL endpoint. 
Contrary to an HTTP server, providing and maintaining a robust and reliable endpoint requires a significant effort that not all publishers are willing or able to make. An alternative query evaluation method is through link traversal, where a query is answered by dereferencing online web resources (URIs) at real time. While several approaches for such a lookup-based query evaluation method have been proposed, there exists no analysis of the types (patterns) of queries that can be directly answered on the live Web, without accessing local or remote endpoints and without a-priori knowledge of available data sources. 
In this paper, we first provide a method for checking if a SPARQL query (to be evaluated on a SPARQL endpoint) can be answered through zero-knowledge link traversal (without accessing the endpoint), and analyse a large corpus of real SPARQL query logs for finding the frequency and distribution of answerable and non-answerable query patterns. Subsequently, we provide an algorithm for transforming answerable queries to SPARQL-LD queries that bypass the endpoints. We report experimental results about  the efficiency of the transformed  queries and discuss the benefits and the limitations of this query evaluation method. 
\end{abstract}

%

\ccsdesc[300]{Information systems~Query languages}

\keywords{SPARQL; Link Traversal; Linked Data; Web of Data; SPARQL-LD}

\maketitle

\section{Introduction}

The Linked Data principles \cite{heath2011linked} has enabled the extension of the Web with a global data space based on open standards and protocols (the so-called \textit{Web of Data}). The current most common way to query this constantly increasing body of knowledge is through SPARQL, where clients send queries to local or remote servers through SPARQL endpoints \cite{feigenbaum2013sparql}. 

However, the low reliability of SPARQL endpoints is the major bottleneck that deters the exploitation of these knowledge bases by real applications \cite{verborgh2016triple,buil2013sparql}. Publicly available endpoints are not optimised for efficiency and they often do not serve many concurrent requests in order to avoid server overloading. For instance, \cite{buil2013sparql} tested 427 public endpoints and found that their performance can vary by up to 3-4 orders of magnitude, while only 32.2\% of public endpoints can be expected to have (monthly) uptimes of 99-100\%.
In general, SPARQL servers are expensive to host and maintain, while providing a reliable public endpoint is a difficult challenge. 
On the contrary, the Linked Data principles provide a simple publishing method which is based on robust web protocols (HTTP, IRI) and can be easily included in existing publishing workflows (e.g., through content negotiation or RDFa).
Thus, there arises the need of alternative, less demanding methods to query Web data \cite{hartig2012foundations,verborgh2016triple}.

Link traversal, in particular, is a query processing method which relies on the Linked Data principles to answer a query by dereferencing online web resources (URIs) dynamically (at query execution time)  \cite{hartig2013overview,hartig2012foundations}. Inspired by this line of research, in this paper we study the query types that can be directly answered through link traversal, without accessing local or remote endpoints and without considering a starting graph or seed URIs for starting the link traversal. Such a \textit{zero-knowledge} query evaluation method is in line with the dynamic nature of the Web, motivates decentralisation, and enables answering queries without requiring data providers to setup and maintain costly endpoints.

\begin{figure}[t]
	\centering
	\includegraphics[width=3.3in]{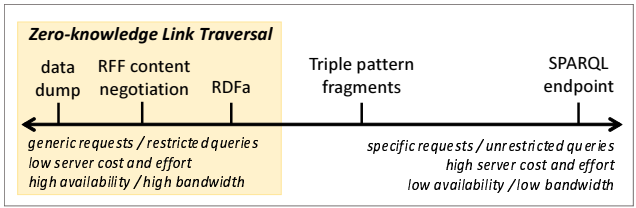}
	\vspace{-2.5mm}
	\caption{Interfaces that allow querying Web data and positioning of zero-knowledge link traversal (this figure is a variation of the figure in \cite{verborgh2014querying}).}
	\label{fig:positioning} 
	\vspace{-4mm}
\end{figure}

Figure \ref{fig:positioning} positions this query execution method in the axis of the existing interfaces that allow querying Web data.  
Zero-knowledge link traversal offers high data availability and bandwidth, and low cost of server setup and maintenance, however it also limits the supported query capabilities. On the contrary, relying on servers offers almost unrestricted query answering, however the server cost is high and the availability and bandwidth low.  

In this paper, we first provide a method for checking if a query (to be evaluated on a SPARQL endpoint) can be answered without accessing any endpoint. We call this query type {\em Linked Data-answerable Queries (LDaQ)}. 
Then, we analyze a large corpus of real query logs from known SPARQL endpoints and study the patterns and frequency of both LDaQ and non-LDaQ. 
We find that more than 85\% of the examined queries are potentially LDaQ, while the majority of them ($>$84\%) follow a few patterns ($\leq$10). 
Then, by exploiting SPARQL-LD \cite{fafalios2015sparql}, a SPARQL 1.1 extension that enables querying any HTTP resource containing RDF data, we provide an algorithm for transforming LDaQ to SPARQL-LD queries that bypass the endpoints. 
We experimentally evaluate the efficiency of the transformed queries and discuss the limitations of this query execution method. We find that more than half of the examined queries can be answered in $<1$ sec, however for queries with large number of intermediate bindings the query execution time can become prohibitively high, thus calling for optimisation methods. 

The implementation of all algorithms and methods described in this paper, as well as the derived data (LDaQ and non-LDaQ patterns), are publicly available.\footnote{\label{foot:git}\url{https://github.com/fafalios/LDaQ}}  

The rest of this paper is organised as follows: 
Section \ref{sec:rw} presents the related literature. 
Section \ref{sec:theproblem} motivates our work and describes the problem. 
Section \ref{sec:methods} introduces the methods for finding LDaQ and transforming them to SPARQL-LD queries.
Section \ref{sec:experiments} presents experimental results.
Finally, Section \ref{sec:conclusion} concludes the paper and discusses interesting directions for future research.

\section{Related Work}
\label{sec:rw}

There are three main paradigms for querying distributed RDF data provided by different Web sources:
i) data centralisation, ii) query federation, and iii) link traversal. 

\subsection{Data Centralisation}

The idea of {\em data centralisation} is to provide a query service
over a collection of RDF data gathered from different sources \cite{sakr2018centralized, tzitzik2014matware}. 
The current de-facto way for querying such repositories is through SPARQL.
Although data centralisation can provide fast responses, it does not exploit the dynamic nature of Web data (the query results may not reflect the more recent data), and it comes at the cost of setting up and maintaining a centralised repository.

A different approach has been proposed in \cite{verborgh2016triple,verborgh2014querying} where the authors introduced {\em Triple Pattern Fragments}, a publishing framework that allows efficient offloading of SPARQL query execution from servers to clients. 
This framework enables servers to maintain high availability rates, allowing querying to scale reliably to much larger numbers of clients. On the downside, the framework requires the setup and maintenance of dedicated servers and clients. 

Contrary to this line of research, in this paper we focus on \textit{zero-knowledge} query execution methods that consider the full potential of the Web and treat a query in isolation, i.e., the input is only a SPARQL query and no other information on how to answer the query is provided (like the URI of an endpoint or resource). 

\subsection{Query Federation}

The idea of {\em query federation} is to provide integrated access to distributed RDF sources on the Web.
DARQ \cite{quilitz2008querying} and SemWIQ \cite{langegger2008semantic} are two
of the first systems to support query federation for SPARQL.
Such systems use a mediator service that transparently distributes the execution of queries to multiple endpoints.
\cite{saleem2016fine} provides a comprehensive analysis and comparison of a large number of endpoint federation systems.
Given the need to address query federation, the SPARQL W3C working group proposed a query federation extension for SPARQL 1.1 \cite{buil2013federating}. The extension defines the {\tt SERVICE} operator which can be used for executing a graph pattern to a remote endpoint. 
Similar to data centralisation, query federation requires the data to be available through SPARQL endpoints. 

SPARQL-LD \cite{fafalios2015sparql,fafaliosquerying} is a generalisation of SPARQL 1.1 which extends the applicability of the {\tt SERVICE} operator to enable querying any HTTP web source containing RDF data, like online RDF files or web pages embedded with RDFa, JSON-LD, or Microformats. 
An important characteristic of SPARQL-LD is that it does not require the named graphs to have been declared, thus one can query datasets returned by a portion of the query, i.e., whose URI derives at query execution time.
\cite{yannakis2018heur} proposes a set of heuristics-based query reordering methods for optimizing the execution of federated queries in both SPARQL 1.1 and SPARQL-LD. 

In this paper we make use of SPARQL-LD for transforming a query (to be evaluated on an endpoint) to a SPARQL-LD query that bypasses the endpoint.

\subsection{Link Traversal}
\label{subsec:linktraversal}

{\em Link traversal} exploits the Linked Data principles \cite{heath2011linked} to dynamically discover data relevant for answering a query \cite{hartig2013overview}. 
The work in \cite{hartig2009executing,hartig2012sparql} follows RDF links between data sources based on URIs in the query and in partial results. The URIs are resolved over the HTTP protocol into RDF data which is continuously added to the queried dataset using an iterator-based pipeline. 
Diamond \cite{miranker2012diamond} is a similar in spirit query engine to evaluate SPARQL queries through link traversal. 
\cite{hartig2011zero} studies how the evaluation order in link traversal affects the size of the results and the query execution cost, and proposes a heuristics-based method to optimize query execution. 
\cite{bouquet2009querying}, \cite{hartig2012sparql} and \cite{harth2012completeness} discuss the notion of {\em completeness} and propose semantics to restrict the range of link traversal queries. 
Finally, index-based approaches rely on pre-built indexes for finding sources to look up during query execution \cite{harth2010data,tian2011enhancing,wagner2012top}.

Regarding more recent works, \cite{umbrich2015link} studies the effectiveness of link traversal-based query execution and proposes reasoning extensions to help finding additional answers.
\cite{fionda2015n} introduces a declarative navigational language for Linked Data, called NautiLOD, which enables to specify data sources by combining navigation and querying capabilities. 
Finally, LDQL \cite{hartig2016ldql} is a more expressive declarative language to query Linked Data through link traversal. 

In our work, we focus on zero-knowledge link traversal and study the types of queries that can by directly answered on the live Web of Data by looking up URIs. The starting point of link traversal is {\em only} the URI(s) that exist in the query's graph pattern and additional URIs are dereferenced only if this is needed for satisfying a triple pattern, i.e. for binding its variables.
This means that, in our case, if the query does not contain URIs, it cannot be evaluated through link traversal. 
This query evaluation method corresponds to the {\em query-reachable} completeness class as introduced in \cite{harth2012completeness}. 

To our knowledge, our work is the first that i) analyses real query logs from known endpoints for finding popular patterns of queries that can be answered or cannot be answered through zero-knowledge link traversal, and ii) provides open source methods to detect answerable queries and transform them to SPARQL-LD queries that are evaluated without accessing endpoints or indexes.  
While recent works have conducted extensive analytical studies on the syntactical and structural characteristics of real SPARQL queries \cite{bonifati2017analytical,saleem2015lsq,bonifati2018darql}, no previous work has analysed queries in terms of their answerability through link traversal.

\section{Motivation \& Problem Description}
\label{sec:theproblem}

Our objective is to study the type of SPARQL queries that can be directly executed on the live Web of Data, without a priori knowledge of available data sources.
The motivation for this zero-knowledge (or query-reachable) approach is threefold: 

\begin{itemize}
	\item The dynamic nature of the Web of Data which constitutes a huge and constantly evolving information space, meaning that we may always need to query a new (unknown) resource not existing in our repository, index or seed URIs. 
	\item The capability to easily run queries directly on the Web of Data, from any client that supports SPARQL, without the need to build and maintain indexes and without requiring data providers to setup and maintain costly servers.
	\item To encourage decentralisation: the Web of Data is increasingly becoming a centralised data space relying on server-side infrastructures \cite{verborgh2017proceedings}. Enabling the execution of SPARQL queries directly on the Web of Data can motivate more publishers to put their data online (e.g., by uploading RDF files), since their data becomes queryable and exploitable without putting effort on setting up and maintaining reliable servers.
\end{itemize}

Consider, for example, the query in Figure \ref{q:example1} which requests the birth date of Barack Obama, and the SPARQL endpoint of DBpedia which can provide an answer to this query.
Our aim is to answer the query without accessing DBpedia's endpoint. 
One approach is to access the URI of Barack Obama used in the query, retrieve the triples contained in this URI, and then run the corresponding triple pattern on these triples. 
Figure \ref{q:example1-sparql-ld} shows a SPARQL-LD query that achieves this. The query uses the extended {\tt SER\-VICE} operator of SPARQL-LD to retrieve and query the triples that are contained in the URI of Barack Obama, thereby bypassing DBpedia's endpoint. 
However, to apply such a transformation, the URI must be dereferenceable and return all the outgoing properties of the corresponding entity. 

\begin{figure}[th]
\vspace{-2mm}
\centering
\footnotesize
\begin{Verbatim}[frame=lines,numbers=left,numbersep=1pt]
SELECT ?birthDate WHERE {
  dbr:Barack_Obama dbo:birthDate ?birthDate }
\end{Verbatim}
\vspace{-3.8mm}
\caption{Example of a LDaQ requesting the birth date of Barack Obama.}
\label{q:example1}

\vspace{0mm}

\centering
\footnotesize
\begin{Verbatim}[frame=lines,numbers=left,numbersep=1pt]
SELECT * WHERE {
  SERVICE <http://dbpedia.org/resource/Barack_Obama> {
    dbr:Barack_Obama dbo:birthDate ?birthDate } }
\end{Verbatim}
\vspace{-3.8mm}
\caption{The transformed SPARQL-LD query of the query in Figure \ref{q:example1}.}
\label{q:example1-sparql-ld}
\vspace{-2mm}
\end{figure}

The query in Figure \ref{q:example2} requests the birth date of all basketball players in DBpedia. In this case, to be able to bypass DBpedia's endpoint, the URI of the DBpedia class {\em Basketball Player} must contain all its incoming properties, i.e. the instances of the class {\em Basketball Player}. 
The query in Figure \ref{q:example2-sparql-ld} shows the corresponding SPARQL-LD query. The query first accesses the URI of the DBpedia class {\em Basketball Player} to retrieve its instances, and then accesses the URI of each instance to retrieve the values of the birth date property.

\begin{figure}[th]
\vspace{-2mm}
\centering
\footnotesize
\begin{Verbatim}[frame=lines,numbers=left,numbersep=1pt]
SELECT ?player ?birthDate WHERE {
  ?player rdf:type dbo:BasketballPlayer ; dbo:birthDate ?birthDate }
\end{Verbatim}
\vspace{-3.8mm}
\caption{Example of a LDaQ requesting the birth date of all basketball players in DBpedia.}
\label{q:example2}

\vspace{0mm}

\centering
\footnotesize
\begin{Verbatim}[frame=lines,numbers=left,numbersep=1pt]
SELECT ?player ?birthDate WHERE {
  SERVICE <http://dbpedia.org/ontology/BasketballPlayer> {
    ?player rdf:type dbo:BasketballPlayer }
  SERVICE ?player { ?player dbo:birthDate ?birthDate } }
\end{Verbatim}
\vspace{-3.8mm}
\caption{The transformed SPARQL-LD query of the query in Figure \ref{q:example2}.}
\label{q:example2-sparql-ld}
\vspace{-2mm}
\end{figure}	

However, not all queries can be transformed to SPARQL-LD queries.
Figure \ref{q:example3} shows two such queries. 
The left query requests all things (of unknown type) having the name {\em \q{Michael Jordan}}, while the right requests the total number of triples. 
Notice that the left query could bypass the endpoint if the URI of the {\tt foaf:name} property provided all the triples that contain it as predicate. However, this is not common in Linked Data and also impractical for large datasets and popular properties (like {\tt rdf:type} and {\tt rdfs:label}). 

\begin{figure}[th]
\vspace{-2.0mm}
\centering
\footnotesize
\begin{minipage}[b]{0.27\textwidth}
\begin{Verbatim}[frame=lines,numbers=left,numbersep=1pt]
SELECT ?entity WHERE {
  ?entity foaf:name "Michael Jordan" }
\end{Verbatim}
\end{minipage} \hspace{2mm}
\begin{minipage}[b]{0.17\textwidth}
\begin{Verbatim}[frame=lines,numbers=left,numbersep=1pt]
SELECT COUNT(*) WHERE {
  ?s ?p ?o }
\end{Verbatim}
\end{minipage}
\vspace{-3.0mm}
\caption{Example of non-LDaQ.}
\label{q:example3}
\vspace{-3.0mm}
\end{figure}

We now define two simple requirements that can enable this functionality for a large portion of SPARQL queries: 
\begin{itemize}
	\item {\bf R1:} URIs must be dereferenceable and return RDF data.  
	\item {\bf R2:} URIs must provide both the incoming and outgoing properties of the corresponding resource (all triples where the URI is the {\em subject} or {\em object}). This includes URIs that represent RDFS/OWL classes, meaning that the URI of a class should return all its instances. 
\end{itemize}

These requirements are in line with the Linked Data principles \cite{heath2011linked}.
An obvious problem of R2 is when the URI represents classes, since the number of instances can be very large for generic classes (like {\em Person} or {\em Location}). We discuss this case at Section \ref{subsec:efficiency}.

\section{Finding \& Transforming Linked Data-answerable Queries}
\label{sec:methods}

In this section, we define the notion of Linked Data-answerable query (Section \ref{subsec:defs}), provide algorithms for checking if a graph pattern or query is Linked Data-answerable (Section \ref{subsec:isLDa}), introduce a method to transform answerable queries to SPARQL-LD queries that bypass the endpoints (Section \ref{subsec:transfrom}), and finally discuss problems and limitations (Section \ref{subsec:problems}). 

The implementation of all algorithms described in this section is publicly available as open source (see Footnote \ref{foot:git}).

\subsection{Linked Data-answerable Graph Patterns and Queries}
\label{subsec:defs}
Following the definitions of \cite{perez2009semantics}, let first $\mathcal{U}$ be an infinite set of URIs, $\mathcal{B}$ an infinite set of blank nodes and $\mathcal{L}$ an infinite set of literals. The union of these sets constitutes the set of \textit{RDF terms}.
A triple $(s, p, o) \in (\mathcal{U} \cup \mathcal{B}) \times \mathcal{U} \times (\mathcal{U} \cup \mathcal{B} \cup \mathcal{L})$
is called an {\em RDF triple}, where $s$ is the {\em subject}, $p$ is the {\em predicate} and $o$ is the {\em object}. We denote by $s(t)$, $p(t)$ and $o(t)$, the subject, predicate and object, respectively, of a triple $t$.
Let also $\mathcal{V}$ be a set of variables that can bind to RDF terms from $\mathcal{U} \cup \mathcal{B} \cup \mathcal{L}$.
A triple $p \in (\mathcal{U} \cup \mathcal{V}) \times (\mathcal{U} \cup \mathcal{V}) \times (\mathcal{U} \cup \mathcal{L} \cup \mathcal{V})$ is called \textit{triple pattern}, while a\textit{ Basic Graph Pattern (BGP)} is a set of triple patterns.
Finally, let $\mathcal{V}^b_i$ be the set of bound variables {\em before} the execution of the $i$-th triple pattern of a BGP. 

We now define the notion of \textit{Linked Data-answerable BGP}: 

\begin{definition}
\vspace{-1mm}
	\label{def:LDaBGP}
	A BGP is Linked Data-answe\-ra\-ble, for short LDaBGP, if its triple patterns $T$ can be brought into an order such that each triple contains at least one URI or bound variable, i.e.:
	$\forall t_i \in T,~ s(t_i) \in \mathcal{U} ~\vee~ o(t_i) \in \mathcal{U} ~\vee~ s(t_i) \in \mathcal{V}^b_i ~\vee~ o(t_i) \in \mathcal{V}^b_i$.	
\vspace{-1mm}	
\end{definition}

This definition corresponds to the \textit{query-reachable} completeness class and the \textit{completely-answerable} BGPs as introduced in \cite{harth2012completeness}. 

Queries containing one or more UNION groups need special handling. Through this operator, SPARQL provides a means of combining graph patterns so that one of several alternative graph patterns may match. 
Consider, for example, the query in Figure \ref{q:exampleUNIONpat} which requests the birth date and place of basketball and football players.
The query contains two UNION groups, each one containing two UNION graph patterns. 
Moreover, the query contains two triples that are not part of the UNION groups (line 4).
To decide if such a query is Linked Data-answe\-ra\-ble, we must check all the graph patterns of each UNION group as well as the triples outside the UNION patterns. However, we should not check them in isolation. For example, the graph patterns of the last UNION group are not Linked Data answerable by themselves, but they are answerable if we consider the bindings of the preceding triples and UNION groups.
We first define the notion of Linked Data-answerable UNION group: 

\begin{definition}
\vspace{-1mm}
	\label{def:LDaUNION}
	A UNION group of BGPs is Linked Data-answe\-ra\-ble if each of its BGPs is Linked Data-answerable. 
\vspace{-1mm}	
\end{definition}

\begin{figure}[th]
\vspace{-2mm}
\centering
\footnotesize
\begin{Verbatim}[frame=lines,numbers=left,numbersep=1pt]
SELECT ?player ?birthDate ?birthPlaceName WHERE {
  { ?player rdf:type dbo:BasketballPlayer } 
  UNION { ?player rdf:type dbo:FootballPlayer }
  ?player dbo:birthDate ?birthDate ; dbo:birthPlace ?place .
  { ?place foaf:name ?birthPlaceName } 
  UNION { ?place rdfs:label ?birthPlaceName } }
\end{Verbatim}
\vspace{-3.8mm}
\caption{Example of a SPARQL query containing two UNION groups.}
\label{q:exampleUNIONpat}
\vspace{-2mm}
\end{figure}

Now we define the notion of \textit{Linked Data-answerable Query} which contains as elements BGPs and UNION groups.

\begin{definition}
\vspace{-1mm}
	\label{def:LDaQ}
	A SPARQL query containing as elements BGPs (sets of triple patterns) and UNION groups  is Linked Data-answerable, for short LDaQ, if its elements can be brought into an order such that each of them is answerable given the variable bindings before the execution of the corresponding BGP / UNION group.
\vspace{-1mm}	
\end{definition}

\subsection{Checking the Answerability of Graph Patterns and Queries}
\label{subsec:isLDa}

Algorithm \ref{alg:isAnswerableBGP} provides a method to find out if a basic graph pattern is Linked Data-answerable or not. 
In brief, the algorithm goes through the triple patterns and finds \q{bindable} variables, i.e., variables that can get bound by dereferencing a URI that exists in the triple, or that can get bound through bindings of other variables. 
If there is at least one non-bindable variable, then the query is not a LDaQ.
The algorithm can also be provided with two optional parameters. The parameter $B$ (bound variables) enables to provide a set of already-bound variables, which is useful for cases where the input graph pattern is part of a query. 
The parameter {\em inUnion} allows specifying that the input graph pattern is part of a UNION group,
thus its bindings must not be considered when checking the other UNION graph patterns of the same UNION group.

\begin{algorithm}[t]
	\scriptsize
	\caption{isLDaBGP}
	\label{alg:isAnswerableBGP}
	\begin{algorithmic}[1]
		\Require $P$: graph pattern, $inUnion$: boolean (optional), $B$: bound variables (optional)  
		\Ensure true or false
		\State $LB = \{\}$ \Comment{Locally-bound variables}
		\State  $M = \{\}$  \Comment{Map a variable to other variables that can help binding it}
		\If{B != null} $LB.addAll(B)$
		\EndIf
		\For{$t : triples(P)$}
		\If{isURI(t.subject)}
		\If{isVariable(t.object)}
		LB.add(t.object) \Comment{Object variable can be bound} \EndIf
		\If{isVariable(t.predicate)}
		LB.add(t.predicate) \Comment{Predicate variable can be bound} \EndIf
		\ElsIf{isURI(t.object)}
		\If{isVariable(t.subject)}
		LB.add(t.subject) \Comment{Subject variable can be bound} \EndIf
		\If{isVariable(t.predicate)}
		LB.add(t.predicate) \Comment{Predicate variable can be bound} 
		\EndIf
		\ElsIf{isVariable(t.subject) \& isVariable(t.object)} 
		\If{(LB.contains(t.subject)} 
		LB.add(T.object) 
		\If{(isVariable(t.predicate)} 
		LB.add(T.predicate) 
		\EndIf
		\ElsIf{(LB.contains(t.object)} 
		LB.add(T.subject) 
		\If{(isVariable(t.predicate)} 
		LB.add(T.predicate)
		\EndIf
		\Else  
		\State M.add(t.subject, t.object) \Comment{Binding of object variable can bind the subject variable}
		\State M.add(t.object, t.subject) \Comment{Binding of subject variable can bind the object variable}
		\If{isVariable(t.predicate)} 
		M.add(t.predicate, \{t.subject, t.object\})  
		\EndIf
		\EndIf
		\EndIf
		\EndFor
		\State $V$ = getAllVariables($P$) \Comment{Set containing all graph pattern variables}
		\For{$v : V$}  \Comment{Check for any variable that cannot be bound}
		\If{$v \notin LB$} 
		\If{$!isBindable(M.get(v))$} \Return false \Comment{Recursively check if the variable can be bound through the bindings of other variables}
		\EndIf
		\EndIf
		\EndFor
		\If{$!inUnion$} $B.addAll(LB);$ \Comment{Add to $B$ the locally bound variables}
		\EndIf
		\State \Return true
	\end{algorithmic}
\end{algorithm}

Algorithm \ref{alg:isAnswerableQuery} checks if a query is Linked Data-answerable or not.
Each triple and UNION group in the query is considered a different {\em element}. 
The algorithm first goes through all the query's elements and checks their answerability using Algorithm \ref{alg:isAnswerableBGP}.
In case the element is a UNION group the algorithm checks the answerability of each UNION's graph pattern. 
If the element is not answerable, it is added to a list of pending elements (since they may be answerable when some variables in another element get bound). If the element is answerable, the list of bound variables is updated with the element's variables.
Then, the algorithm checks the pending elements. In each loop, at least one new element must get answerable, otherwise the query is not Linked Data-answerable.

\begin{algorithm}[t]
	\scriptsize
	\caption{isLDaQ}
	\label{alg:isAnswerableQuery}
	\begin{algorithmic}[1]
		\Require $Q$: Query graph pattern 
		\Ensure true or false
		\State $B = \{\}$; $PENDING = \{\}$ \Comment{Bound variables; Pending query elements}
		\State $E$ = getElements($Q$) \Comment{Each triple and each UNION group is considered a different element}
		\For{$e : E$}  \Comment{For each triple or UNION group}  
		\If{$isTripleElement(e)$} \Comment{The element is a triple}  
		\If{$!isLDaBGP(e, false, B)$}
		\State $PENDING.add(e)$ \Comment{Add this element to the list of pending elements}  
		\EndIf
		\Else \Comment{The element is a UNION group} 
		\State $allAnswerable = true$
		\For{$u : getUnionGraphPatterns(e)$}
		\If{$!isLDaBGP(u, true, B)$} $allAnswerable = false$; {\bf break}
		\EndIf
		\EndFor
		\If{$allAnswerable$} $B.addAll(getVariables(e))$ 
		\Else ~$PENDING.add(e)$  \Comment{Add this element to the list of pending elements}  
		\EndIf
		\EndIf
		\EndFor
		\While{$!PENDING.isEmpty()$} \Comment{While there exist pending elements}
		\State $foundNew = false$ \Comment{In each FOR loop we must find a new answerable element}
		\For{$pe : PENDING$} \Comment{For each pending element}  
		\If{$isTripleElement(pe)$}
		\If{$isLDaBGP(pe, false, B)$}
		\State $foundNew = true$; $PENDING.remove(pe)$  
		\EndIf
		\Else \Comment{The pending element is a UNION group} 
		\State $allAnswerable = true$
		\For{$u :  getUnionGraphPatterns(pe)$}
		\If{$!isLDaBGP(u, true, B)$}  $allAnswerable = false$; {\bf break}
		\EndIf
		\EndFor
		\If{$allAnswerable$}
		\State $foundNew = true$; $PENDING.remove(pe)$
		\State $B.addAll(getVariables(pe))$ 
		\EndIf
		\EndIf
		\EndFor
		\If{$!foundNew$} \Return $false$ \Comment{No new answerable element was found}  
		\EndIf
		\EndWhile
		\State \Return true \Comment{The query is Linked Data-answerable}
	\end{algorithmic}
\end{algorithm}

\subsection{Transforming to SPARQL-LD}
\label{subsec:transfrom}

We now provide a method to transform a LDaQ to a SPARQL-LD query that evaluates its graph pattern directly over the live Web of Linked Data without accessing local or remote endpoints. 
Such a {\em transformation-based} approach to run LDaQ offers the ability to directly make use of this query execution method through existing instances of SPARQL-LD, i.e., without the need to setup a dedicated server that supports the execution of link traversal queries.  

Algorithm \ref{alg:transformBGP} transforms a BGP to a SPARQL-LD graph pattern. 
The algorithm goes through the triples and creates SERVICE patterns.
Specifically, if the triple contains a URI or a bound variable, it checks if there is already a SERVICE pattern for the same URI/ variable. If so, the triple is just added to its graph pattern, otherwise a new SERVICE pattern is created.
Notice that if both the subject and object are URIs, we decide to look-up only the subject URI.
If the triple does not contain a URI or bound variable, it is added to a list of pending triples. Since the BGP is Linked Data-answerable, these triples require the binding of another variable (existing in a subsequent triple).  
After checking all triple patterns, the algorithm goes through the pending triples and, correspondingly, creates new SERVICE patterns or updates the existing ones. 

Algorithm \ref{alg:transformQuery} transforms a Linked Data-answerable query (that may also contain UNION groups) to a SPARQL-LD query.
The algorithm goes through the query's elements (which can be either single triples or UNION groups) and checks if they are Linked Data-answerable. If so, the procedure {\tt INCLUDE} is executed. This procedure includes the element to the SPARQL-LD query, either by appending it to an existing SERVICE or by creating a new one.
If the element is not Linked Data-answerable, it is added to a list of pending elements whose transformation requires the binding of a variable existing in a subsequent triple or UNION group. 
Then the algorithm goes through the pending elements and includes them in the transformed SPARQL-LD query once they get answerable.

\begin{algorithm}
	\scriptsize
	\caption{transformBGP}
	\label{alg:transformBGP}
	\begin{algorithmic}[1]
		\Require Basic graph pattern $P$
		\Ensure SPARQL-LD query pattern $P'$
		\State $P' = \{\}$ \Comment{SPARQL-LD graph pattern}
		\State $B = \{\}$; $PENDING = \{\}$ \Comment{Bound variables; Pending triple patterns}
		\For{$t \in getTriples(P)$} \Comment{For each triple pattern}
		\If{$isURI(t.subject) ~||~ isURI(t.object)$} 
		\State $u = isURI(t.subject)~?~t.subject~:~t.object$ \Comment{Consider subject or object URI}
		\If{$P'.containsService(u)$} \Comment{There is a service pattern for the same URI}
		\State $P'.getService(u).add(t)$  \Comment{Add the triple pattern to its graph pattern}
		\Else ~ $P'.add(new ServicePattern(u, t))$ \Comment{Create a new service pattern}
		\EndIf
		\State $updateBoundVariables(t, B)$ \Comment{Update the set of bound variables}
		\ElsIf{$B.contains(t.subject) ~||~ B.contains(t.object)$}
		\State $v = B.contains(t.subject)~?~t.subject ~ t.object$ \Comment{Consider the subject or object variable}
		\If{$P'.containsService(v)$} \Comment{There is a service pattern for the same variable}
		\State $P'.getService(v).add(t)$ \Comment{Add the triple pattern to its graph pattern}
		\Else ~$P'.add(new ServicePattern(v, t))$ \Comment{Create a new service pattern}
		\EndIf
		\State $updateBoundVariables(t, B)$ 
		\Else ~$PENDING.add(t)$ \Comment{Transform this triple pattern later}
		\EndIf
		\EndFor
		\While{$!PENDING.isEmpty()$} 
		\For{$pt \in PENDING$} \Comment{For each pending triple pattern}
		\If{$B.contains(pt.subject) ~||~ B.contains(pt.object)$}
		\State $v = B.contains(pt.subject)~?~t.subject ~ t.object$ 
		\If{$P'.containsService(v)$} $P'.getService(v).add(pt)$ 
		\Else ~ $P'.add(new ServicePattern(v, pt))$ 
		\EndIf
		\State $updateBoundVariables(pt, B)$; $PENDING.remove(pt)$ 
		\EndIf
		\EndFor
		\EndWhile
		\State \Return $P'$
	\end{algorithmic}
\end{algorithm}

\begin{algorithm}
	\scriptsize
	\caption{transformQuery}
	\label{alg:transformQuery}
	\begin{algorithmic}[1]
		\Require $Q$: query graph pattern
		\Ensure $Q'$: SPARQL-LD query pattern
		\State $Q' = \{\}$; $B = \{\}$; $PENDING = \{\}$ \Comment{SPARQL-LD pattern; Bound vars; Pending elements}
		\State $E$ = getElements($Q$) \Comment{Each triple and each UNION group is considered a different element}
		\For{$e : E$}  \Comment{For each triple or UNION group}  
		\If{$isTripleElement(e)$} \Comment{The element is a triple}  
		\If{$isLDaBGP(e, false, B)$} {\tt INCLUDE}$(Q', e, B)$
		\Else   
		~$PENDING.add(e)$ \Comment{Transform it later}  
		\EndIf
		\Else \Comment{The element is a UNION group} 
		\State $allAnswerable = true$
		\For{$u : getUnionGraphPatterns(e)$}
		\If{$!isLDaBGP(u, true, B)$}  ~$allAnswerable = false$; {\bf break}
		\EndIf
		\EndFor
		\If{$allAnswerable$}  $B.addAll(getVariables(e))$; {\tt INCLUDE}$(Q', e, B)$ 
		\Else
		~$PENDING.add(e)$  \Comment{Transform it later}  
		\EndIf
		\EndIf
		\EndFor
		\While{$!PENDING.isEmpty()$} \Comment{While there exist pending elements}
		\For{$pe : PENDING$} \Comment{For each pending element}  
		\If{$isTripleElement(pe)$}
		\If{$isLDaBGP(pe, false, B)$}
		\State {\tt INCLUDE}$(Q', e, B)$; $PENDING.remove(pe)$  
		\EndIf
		\Else \Comment{The pending element is a UNION group} 
		\State $allAnswerable = true$
		\For{$u :  getUnionGraphPatterns(pe)$}
		\If{$!isLDaBGP(u, true, B)$}  $allAnswerable = false$; {\bf break}
		\EndIf
		\EndFor
		\If{$allAnswerable$}
		\State {\tt INCLUDE}$(Q', e, B)$
		\State $PENDING.remove(pe)$; $B.addAll(getVariables(pe))$ 
		\EndIf
		\EndIf
		\EndFor
		\EndWhile
		\State \Return $Q'$
	\end{algorithmic}
\end{algorithm}

\subsection{Problems and Limitations}
\label{subsec:problems}

There are some data access issues that must be taken into account when running queries over the live Web of Data \cite{hartig2013overview}. 
In brief, dereferencing a URI may result in the retrieval of an unforeseeable large set of RDF triples, while some servers put restrictions on clients such as serving only a limited number of requests per second. Thus, a link traversal-based query execution system should implement a politeness policy to avoid overloading servers, e.g., by respecting the {\em robots.txt} protocol that allows web sites to demand delays between subsequent requests from the same client. 

In this paper we do not examine the case of DESCRIBE queries, as well as of queries containing the operators FROM, FROM NAMED / GRAPH, and SERVICE. These queries correspond to around 15\% of the queries submitted to popular SPARQL endpoints \cite{bonifati2017analytical}. 
Indicatively, for DESCRIBE queries we can just look up the provided URI and return all its  triples. 
For FROM and FROM NAMED / GRAPH queries, the triples of the provided resource should be fetched and the corresponding graph pattern can be directly executed over these triples (without checking its answerability). Finally, SERVICE patterns over remote endpoints can be also transformed to SPARQL-LD queries if their graph pattern is Linked Data-answerable.
We leave the implementation of all these cases as part of our future work.

\section{Experimental Results}
\label{sec:experiments}

\subsection{Datasets}

We experimented with real SPARQL query logs provided by the Linked SPARQL Queries Dataset (LSQ) \cite{saleem2015lsq} and the USEWOD series of workshops \cite{luczak2016usewod}. From LSQ, we used all the queries of Linked Geo Data (LGD), Semantic Web Dog Food (SWDF), British Museum (BM), and DBpedia, while from USEWOD we used the queries of 
LGD, SWDF, BIO2RDF, and the more recent DBpedia 2014 and 2015 queries. 
The total number of queries in these datasets is 67,849,121.

We first fixed some common errors found in the queries (like the absence of popular prefixes), and then used Jena 3.2 to parse them and get their graph pattern. In our experiments, we did not consider the queries that are not valid according to Jena 3.2 and the queries that use property paths or contain one of the following operators: DESCRIBE, FROM, GRAPH, SERVICE, MINUS, EXISTS, BIND, VALUES, SUB-SELECT (nested queries). 

Table \ref{tab:datasetStats} shows the main statistics per dataset. The last column shows the total number of \textit{unique} queries that we consider in our analysis. For finding the unique queries, we compared only the query graph patterns, i.e., without considering the prefixes, the SELECT clause, and any ORDER/GROUP BY operators.

\begin{table}
    \vspace{-2mm}
	\centering
	\caption{Dataset statistics.}
	\label{tab:datasetStats}
	\vspace{-3.5mm}
	\renewcommand{\arraystretch}{0.65}
    \setlength{\tabcolsep}{3.5pt}
	\footnotesize
	\begin{tabular}{lrrrrrrr}
		\toprule
		\makecell{Dataset} &
		\makecell{\#Queries} &
		\makecell{\#Invalid} &
		\makecell{\#Unconsidered} &
		\makecell{\#Remaining} & 
		\makecell{\#Unique}\\
		\midrule
		LGD        & 4,240,736     & 456,393   & 1,148,809   & 2,635,534    & 670,809    \\
		SWDF       & 13,990,138    & 224,849   & 3,326,767   & 10,438,522   & 789,049   \\
		BM         & 129,989       & 0         & 0           & 129,989      & 129,989    \\
		BIO2RDF    & 192,057       & 47        & 2           & 192,008      & 62,819     \\
		DBPEDIA    & 49,296,201    & 2,003,381 & 3,869,723   & 43,423,097   & 16,028,271 \\
		\bottomrule
	\end{tabular}
	\vspace{-2mm}
\end{table}

\subsection{Pattern-based analysis of LDaQ and non-LDaQ}

We examined the Linked Data-answerability of all unique queries (using Algorithm \ref{alg:isAnswerableQuery}) as well as the pattern (template) they follow. For getting the pattern of a query, we considered only its graph pattern (text under WHERE), removed the FILTER operators, and replaced all variables, URIs, literals, and blank nodes with the strings [V], [U], [L], and [B], respectively.\footnote{The implementation of pattern extraction is publicly available (cf. Footnote \ref{foot:git}).}
For example, the pattern of the query in Figure \ref{q:exampleUNIONpat} is the following:

\scriptsize
\begin{Verbatim}[frame=lines]
{ [V] [U] [U] } UNION { [V] [U] [U] } [V] [U] [V] ; [U] [V] 
{ [V] [U] [V] } UNION { [V] [U] [V] }
\end{Verbatim}
\normalsize

Table \ref{tab:transfStats} shows the number and percentage of LDaQ and non-LDaQ, and the corresponding number of unique patterns. 
We notice that the percentage of LDaQ is more than 85\% in all datasets. BM and BIO2RDF contain the highest percentage of LDaQ (99,9\% and 96.7\%, respectively), however we also notice that the number of unique patterns in these two datasets is very small (only 5 for BM and 14 for BIO2RDF) which means that, possibly, the queries in these collections come from fixed templates. As regards DBPEDIA, the largest and most popular dataset in our collection, we see that the majority of its unique queries (87.7\%) are potentially LDaQ.

\begin{table}
    \vspace{-2mm}
	\centering
	\caption{Linked Data answerable and no-answerable queries and unique patterns.}
	\label{tab:transfStats}
	\vspace{-3.5mm}
	\renewcommand{\arraystretch}{0.65}
    \setlength{\tabcolsep}{3.0pt}
	\footnotesize
	\begin{tabular}{lrrrrr}
		\toprule
		\makecell{Dataset} &
		\makecell{\#Test\\queries} &
		\makecell{\#LDaQ} &
		\makecell{\#LDaQ\\patterns} &
		\makecell{\#Non-LDaQ} & 
		\makecell{\#non-LDaQ\\patterns} \\
		\midrule
		LGD        & 670,809     & 572,720 (85.4\%)      & 444   & 98,089 (14.6\%)     & 197 \\
		SWDF       & 789,049     & 677,923 (85.9\%)      & 570   & 111,126 (14.1\%)    & 202 \\
		BM         & 129,989     & 129,936 (99.9\%)      & 4     & 53 (0.04\%)         & 1   \\
		BIO2RDF    & 62,819      & 60,740 (96.7\%)       & 9     & 2,079 (3.30\%)      & 5  \\
		DBPEDIA    & 16,028,271  & 14,053,584 (87.7\%)   & 2,816 & 1,974,687 (12.3\%)  & 780 \\
		\bottomrule
	\end{tabular}
	\vspace{-1mm}
\end{table}

\begin{figure}[h]
\vspace{-3mm}
	\centering
	\subfloat[Answerable patterns]{
		\includegraphics[scale=.4]{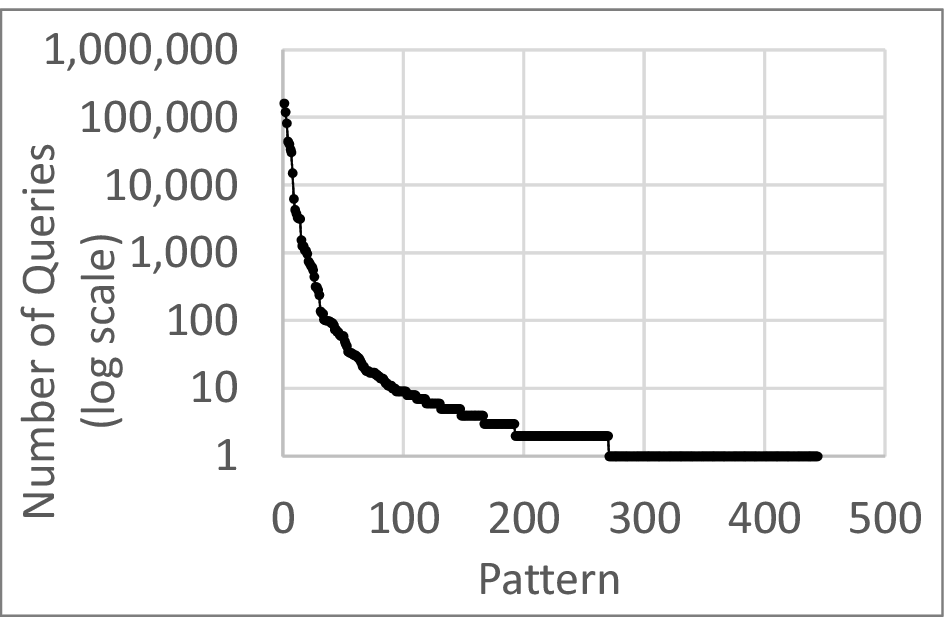}
		\label{fig:transfLGD}
	}
	\subfloat[Not answerable patterns]{
		\includegraphics[scale=.4]{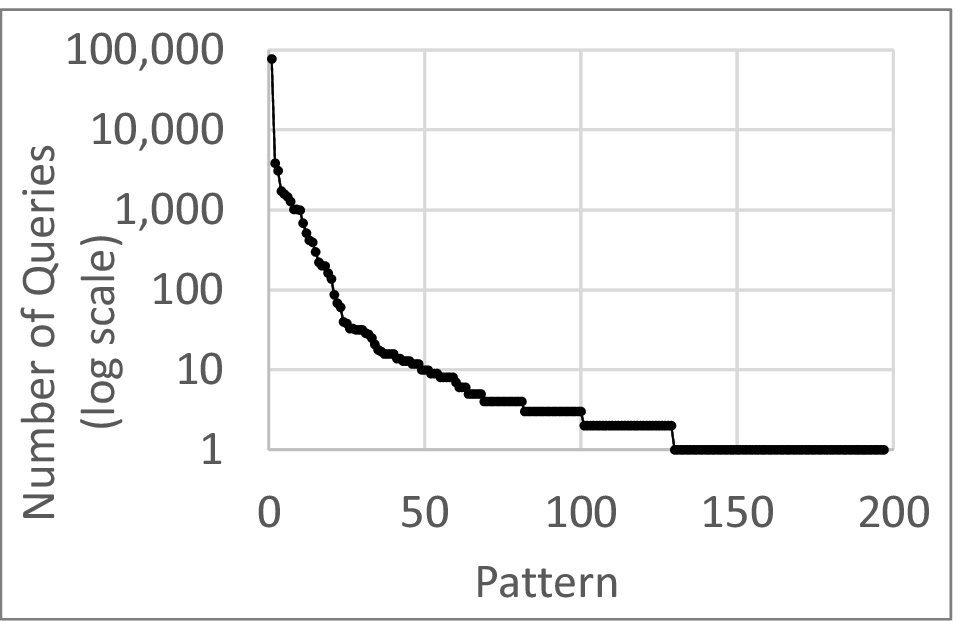}
		\label{fig:noTransfLGD}
	}
	\vspace{-3mm}
	\caption{LDaQ and non-LDaQ pattern distribution in LGD.}
	\label{fig:distrPatternsLGD}

	\centering
	\subfloat[Answerable patterns]{
		\includegraphics[scale=.4]{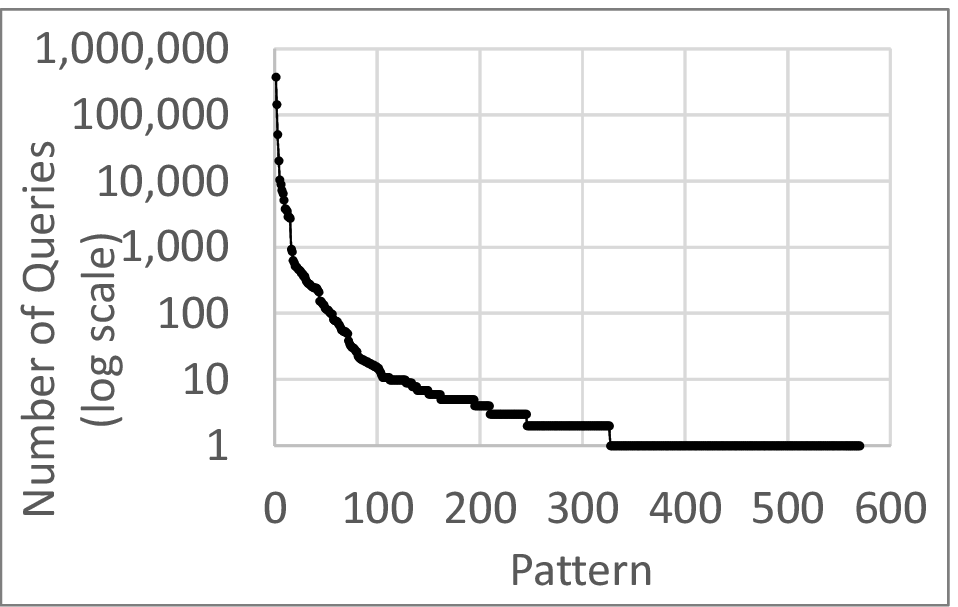}
		\label{fig:transfSWDF}
	}
	\subfloat[Not answerable patterns]{
		\includegraphics[scale=.4]{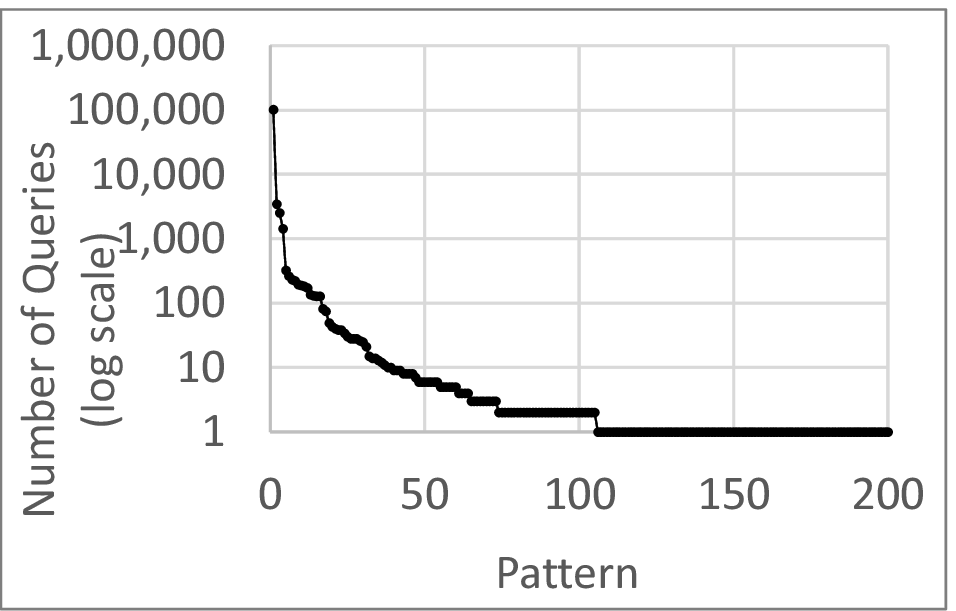}
		\label{fig:noTransfSWDF}
	}
	\vspace{-3mm}
	\caption{LDaQ and non-LDaQ pattern distribution in SWDF.}
	\label{fig:distrPatternsSWDF}

	\centering
	\subfloat[Answerable patterns]{
		\includegraphics[scale=.4]{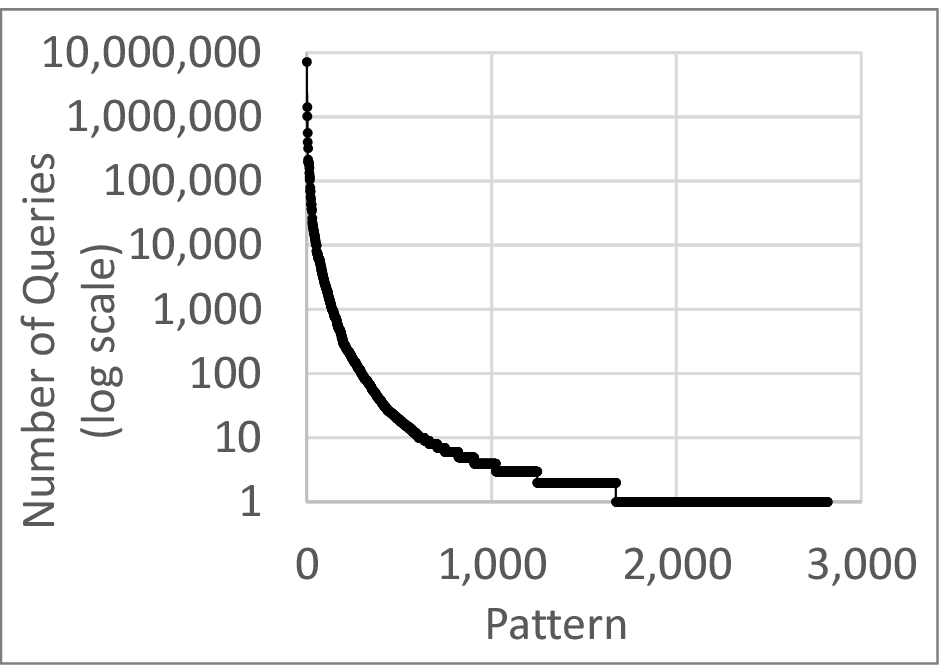}
		\label{fig:transfDBpedia}
	}
	\subfloat[Not answerable patterns]{
		\includegraphics[scale=.4]{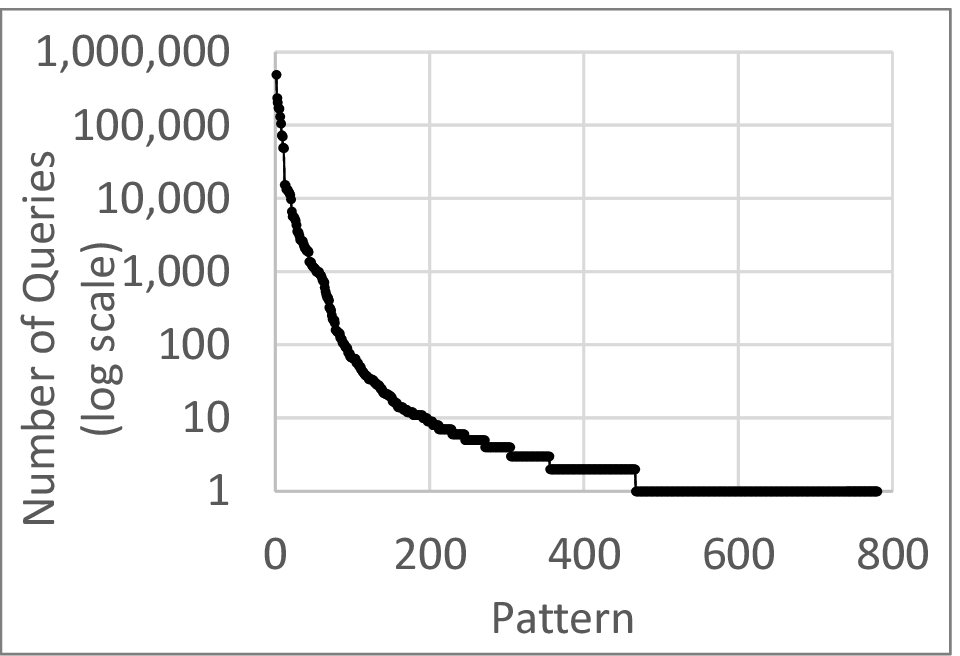}
		\label{fig:noTransfDBpedia}
	}
	\vspace{-3mm}
	\caption{LDaQ and non-LDaQ pattern distribution in DBPEDIA.}
	\label{fig:distrPatternsDBpedia}
	\vspace{-3mm}
\end{figure}

Figures \ref{fig:distrPatternsLGD}-\ref{fig:distrPatternsDBpedia} show the distribution of LDaQ and non-LDaQ for LGD, SWDF, and DBPEDIA.
We notice that all follow a similar power-law distribution: there is a very small number of patterns having  a very large number of queries and a long tail of patterns each one having only a few queries. 
The top-10 LDaQ patterns in LGD and SWDF correspond to the 95\% of all answerable queries, and the top-10 non-LDaQ to the 98\% and 96\%, respectively, of all non-answerable queries. 
Regarding the DBPEDIA dataset, the top-10 LDaQ patterns correspond to the 84\% of the answerable queries and the non-LDaQ to the 86\% of the non-answerable queries.

The listings in Figures \ref{top5LGD}-\ref{top5DBpedia} show the top-5 LDaQ and non-LDaQ patterns for LGD, SWDF and DBPEDIA (where {\tt UN} = UNION, {\tt OPT} = OPTIONAL).
We notice that the majority of the frequent LDaQ patterns are short and request either the properties of a URI (like the patterns {\tt [U] [V] [V]} and {\tt [U] [U] [V]}) or the URIs having a specific property value (like the patterns {\tt [V] [V] [U]} and {\tt [V] [U] [U]}).
Regarding the non-LDaQ queries, we see that {\tt [V] [U] [V]} and {\tt [V] [U] [L]} are the more frequent patterns. 
In DBPEDIA, it is interesting that some of the top patterns are long and contain many UNION and OPTIONAL operators. These patterns probably correspond to a large number of similar (template-based) queries, possibly submitted by a small number of clients. 

The full lists of LDaQ and non-LDaQ patterns are publicly available (cf. Footnote \ref{foot:git}).

\begin{figure}
\centering
\scriptsize
\begin{minipage}[b]{0.48\textwidth}
\begin{Verbatim}[frame=lines]
1 [U] [V] [V]  
2 [V] [V] [U] 
3 OPT { [U] [U] [V] } 
4 [V] [U] [U] ; [V] [V] 
5 [V] [U] [U] 
\end{Verbatim}
\end{minipage}

\vspace{1mm}

\begin{minipage}[b]{0.48\textwidth}
\begin{Verbatim}[frame=lines]
1 [V] [U] [V]
2 [V] [U] [V] . [V] [U] [V]
3 [V] [U] [L]
4 [V] [U] [L] OPT {[V] [U] [V]} OPT {[V] [U] [V]}
5 [V] [V] [V]
\end{Verbatim}
\end{minipage}
\vspace{-5.5mm}
\caption{Top-5 LDaQ (up) and non-LDaQ (down) patterns in LGD.}
\label{top5LGD}

\vspace{1mm}

\begin{minipage}[b]{0.48\textwidth}
\begin{Verbatim}[frame=lines]
1 [U] [U] [V] 
2 [V] [U] [U] 
3 [U] [V] [V] 
4 { [U] [V] [V] } UN { [V] [V] [U] }
5 [U] [V] [V] OPT { [U] [U] [V] }
\end{Verbatim}
\end{minipage}

\vspace{1mm}

\begin{minipage}[b]{0.48\textwidth}
\begin{Verbatim}[frame=lines]
1 [V] [U] [L] 
2 [V] [U] [V] 
3 [V] [U] [L] ; [U] [V] . [V] [U] [V] 
4 OPT { [V] [U] [V] }
5 [V] [U] [V] ; [U] [V] 
\end{Verbatim}
\end{minipage}
\vspace{-5.5mm}
\caption{Top-5 LDaQ (up) and non-LDaQ (down) patterns in SWDF.}
\label{top5SWDF}

\vspace{1mm}

\begin{minipage}[b]{0.48\textwidth}
\begin{Verbatim}[frame=lines]
1 [U] [U] [V]
2 { [U] [U] [U] } UN { [U] [U] [U] }
3 [V] [U] [U] ; [U] [L] . [V] [U] [U] { [V] [U] [V] } UN { [V] [U] [V] } UN { 
  [V] [U] [V] } UN { [V] [U] [V] } { [V] [U] [V] } UN { [V] [U] [V] } OPT { 
  [V] [U] [V] } OPT { [V] [U] [V] } OPT { [V] [U] [V] }
4 [U] [V] [V]
5 { [V] [U] [U] } UN { [V] [U] [U] } [V] [U] [L] . [V] [U] [V] ; [U] [L] ; [U] [V] 
\end{Verbatim}
\end{minipage}

\vspace{1mm}

\begin{minipage}[b]{0.48\textwidth}
\begin{Verbatim}[frame=lines]
1 { [V] [U] [L] } UN { [U] [U] [V] } [V] [U] [V] ; [U] [V] . [V] [U] [V]
2 [V] [U] [L] ; [V] [V] OPT { [V] [U] [V] }
3 { [V] [U] [L] } UN { [V] [U] [V] ; [U] [L] } UN { [V] [U] [V] ; [U] [L] } OPT { 
  [V] [U] [V] } OPT { [V] [U] [V] ; [U] [V] } OPT { [V] [U] [V] } OPT { [V] [U] 
  [V] } OPT { [V] [U] [V] } OPT { [V] [U] [V] } OPT { [V] [U] [V] }
4 [V] [U] [L]
5 [V] [V] [V] . [V] [U] [L]
\end{Verbatim}
\end{minipage}
\vspace{-5.5mm}
\caption{Top-5 LDaQ (up) and non-LDaQ (down) patterns in DBPEDIA.}
\label{top5DBpedia}
\vspace{-3mm}
\end{figure}

\subsection{Efficiency of the transformed queries}
\label{subsec:efficiency}

\subsubsection{Querying a single URI} 
This is the simplest case where we request one or more properties (incoming or outgoing) of a single resource (patterns like {\tt [U] [V] [V]} and  {\tt [V] [U] [U]}). This query type corresponds to around 77\% of all unique queries in the SWDF dataset, 70\% in LGD, 97\% in BIO2RDF, and 56\% in DBpedia.

As shown in \cite{fafaliosquerying}, the time to answer this query type is proportional to the number of triples contained in the resource. Querying a resource of 10,000 triples requires around 1 sec while the time increases to 30 secs for resources with 1M triples. The same work examined the case of querying DBpedia URIs and showed that the average query time is around 320 ms if we access the N3 files and 650 ms through content negotiation, while the time to run the same query at DBpedia's endpoint is around 300 ms.
Requesting one or more of the outgoing properties of a URI corresponds to 52\% of all unique queries in the examined DBPEDIA dataset. This means that more than half of the queries can bypass the endpoint and be efficiently answered through link traversal.
In general, this query type does not increase the data that is transferred over the network, while for queries requesting the outgoing properties of a specific entity, the query execution time is very low (since the number of triples is usually small). The time can be high for queries requesting the incoming properties of resources representing classes since in some cases the number of instances can be very large. For example, in DBpedia 2016, there are 3,218,716 instances of type {\em dbo:Person}.\footnote{\url{http://wiki.dbpedia.org/dbpedia-2016-04-statistics}}
If we consider that querying a resource of 1M triples requires around 30 seconds \cite{fafaliosquerying}, the time to retrieve all instances of such a general class is around 1.5 minute. 
Requesting the incoming properties of a URI corresponds to around 3.6\% of all unique queries in the DBPEDIA dataset, 20\% in SWDF, and 25\% in LGD.

\subsubsection{Querying multiple URIs} 
This case includes the majority of queries containing joins (patterns like {\tt [V] [U] [U] ; [U] [V]}).
For instance, the query in Figure \ref{q:example2-sparql-ld}, which requests a specific property value (birth date) of all entities of a specific type (basketball players), is such a query. 
The query execution time in this case highly depends on the number of intermediate bindings. 

We run experiments for the popular pattern {\tt [V] [U] [U] ; [U] [V]} for different number of intermediate results. 
The submitted query requests the English label of all instances belonging to a particular class. We tested the following Wikicat classes containing varied number of instances: 
(a) American Civil Rights Lawyers (136 instances),
(b) Video Artists (262 instances),
(c) People From Sheffield (502 instances),
(d) American Magazines (1,030 instances), and
(e) American Male Film Actors (9,787 instances).
We run the queries 10 times in different time points, and computed the average time to execute the corresponding SPARQL-LD query and store the results. 
We tried two different methods: i) non-optimised (sequential fetching of remote resources), and ii) optimised (using a simple parallelisation method which runs maximum 10 parallel threads at the same time for fetching the remote resources). 

Table \ref{tab:queryExecTime} shows the results. 
As expected, the query execution time is proportional to the number of intermediate bindings since the query needs to fetch the triples of each binding corresponding to a URI. 
We see that for large number of bindings the query execution time can be very high, especially if we do not optimise the query evaluation process. 
Such queries can highly increase the traffic of the HTTP server, thus the corresponding SPARQL-LD implementation should apply a politeness policy (c.f. Section \ref{subsec:problems}). 

\begin{table}
\vspace{-2mm}
	\centering
	\caption{Query execution time (in seconds) of the transformed SPARQL-LD queries for different number of intermediate bindings (resources to be fetched): (a) 136, (b) 262, (c) 502, (d) 1,030, (e) 9,787.}
	\label{tab:queryExecTime}
	\vspace{-2mm}
	\renewcommand{\arraystretch}{0.65}
    \setlength{\tabcolsep}{3.5pt}
	\small 
	\begin{tabular}{lrrrrrrr}
		\toprule
		\makecell{ } &
		\makecell{(a)} &
		\makecell{(b)} &
		\makecell{(c)} &
		\makecell{(d)} & 
		\makecell{(e)}\\
		\midrule
		Non-optimised   & 26    & 44    & 79   & 152   & 1,322  \\
		Optimised       & 7     & 13    & 24   & 48    & 423    \\
		\bottomrule
	\end{tabular}
	\vspace{-3mm}
\end{table}

\section{Conclusions}
\label{sec:conclusion}

We essentially investigated the case where instead of having heavy-loaded servers (SPARQL endpoints) and light clients (SPARQL clients), we have very light servers (just Linked Data hosting) and heavier clients.
This scenario could be beneficial not only in terms of managerial costs, but also in terms of load balancing and robustness, however it could  increase the data that should be transferred over the network and the overall query execution time.

To this end, we introduced a method for checking whether a SPARQL query can be answered on the live Web of Data without accessing any endpoint.
We analysed a large dataset of real SPARQL query logs for identifying frequent answerable and non-answerable query patterns. The analysis showed that more than 85\% of the examined queries are potentially Linked Data-answerable, while the majority ($>$84\%) of both answerable and non-answerable queries follow a few ($\leq$10) specific patterns.
Subsequently we provided an algorithm for transforming Linked Data-answerable queries to SPARQL-LD queries that bypass the endpoints. Such a method to query Linked Data is based on standard and well-established Web technologies (HTTP, URI) and does not require the installation and maintenance of new servers and clients.

With respect to the efficiency of the transformed queries, the query execution time highly depends on the number of remote resources that need to be accessed and the size of these resources (number of triples). We saw that more than half of the examined DBpedia queries can be answered through this method in $<1$ sec. However, for queries with large number of intermediate bindings, which in turn might require large number of URI lookups, the query execution time can become prohibitively high.

In general, we saw that, as expected, we cannot totally avoid SPARQL endpoints and offer unrestricted query capabilities through zero-knowledge link traversal. 
We also expect that query evaluation is (almost) always faster in endpoints than through link traversal, since endpoints rely on pre-built indexes/databases.
Nevertheless, our results showed that this query evaluation method can be offered efficiently for a large portion of queries, which could potentially decrease the load of these endpoints and increase their availability.

Regarding future work, an interesting direction is the design of adaptive query processing methods that combine different query execution strategies based on the load of the servers, the availability of the remote sources, and the estimated efficiency of query execution. Another interesting direction is the study of approaches to improve the execution time of the transformed SPARQL-LD queries, e.g., through caching or better query planning. 
Further examination of the non-answerable query patterns is also needed. For example, would a different policy for publishing Linked Data be beneficial for making more queries answerable?

\bibliographystyle{ACM-Reference-Format}
\balance
\bibliography{SAC2019__BIB} 

\end{document}